\documentclass{article}
\usepackage[english]{babel}
\usepackage[letterpaper,top=2cm,bottom=2cm,left=1.5cm,right=1.5cm,marginparwidth=1.75cm]{geometry}
\usepackage{authblk}
\usepackage{amsmath}
\usepackage{graphicx}
\usepackage{float}
\usepackage[colorlinks=true, allcolors=blue]{hyperref}
\title{\vspace{-0cm}
  Forecasting Inter-Destination Tourism Flow via a Hybrid Deep Learning Model}

\author[1]{Hanxi Fang} 
\author[2]{Song Gao}
\author[1]{Feng Zhang}
\affil[1]{School of Earth Science, Zhejiang University}
\affil[2]{Geospatial Data Science Lab, University of Wisconsin-Madison}
\date{}

\begin{document}
\maketitle
\begin{abstract}
    Tourists often go to multiple tourism destinations in one trip. The volume of tourism flow between tourism destinations, also referred to as ITF (Inter-Destination Tourism Flow) in this paper, is commonly used for tourism management on tasks like the classification of destinations' roles and visitation pattern mining. However, the ITF is hard to get due to the limitation of data collection techniques and privacy issues. It is difficult to understand how the volume of ITF is influenced by features of the multi-attraction system. To address these challenges, we utilized multi-source datasets and proposed a graph-based hybrid deep learning model to predict the ITF. The model makes use of both the explicit features of individual tourism attractions and the implicit features of the interactions between multiple attractions. Experiments on ITF data extracted from crowdsourced tourists' travel notes about the city of Beijing verified the usefulness of the proposed model. Besides, we analyze how different features of tourism attractions influence the volume of ITF with explainable AI techniques. Results show that popularity, quality and distance are the main three influential factors. Other features like coordinates will also exert an influence in different ways. The predicted ITF data can be further used for various downstream tasks in tourism management. The research also deepens the understanding of tourists' visiting choice in a tourism system consisting of multiple attractions.
\end{abstract}

\section{Introduction}
Many tourists visit multiple tourism destinations in one trip. This often happens on inter-city level or inter-attraction level within a city. For example, people may visit both cities Beijing and Shanghai in their trip to China (inter-city level), and may visit both the Great Wall and the Forbidden City in their trip to the city of Beijing (inter-attraction level). This phenomenon leads to the research of the multi-destination tourism system and the spatial structure of tourism destinations~\cite{xu2022understanding,kim2022visitor,ramadhani2021mapping}. Commonly studied topics in these research include: what is the role of a certain destination in the whole multi-destination system, which destinations are more connected to each others and thus form a community, and how different features of destinations and tourists influence people's visitation patterns, etc. The answers to these questions can be used for designing better marketing strategies and making suggestions for tourism management. 

In these researches, the most commonly-used data is Inter-Destination Tourism Flow (ITF), which is the volume of tourism flow between tourism destinations. Some of the ITF used are directed, which consider the visiting order; Others are undirected, which only consider if two destinations co-occur in a trip. Both the directed ITF and undirected ITF are used widely in studies for multi-destination tourism system. To quantitatively model the interaction of multiple destinations, a multi-destination network or graph is often built. The nodes of the network usually represent destinations, and the edge weight represents the volume of ITF~\cite{ramadhani2021mapping,xu2022understanding,shih2006network}. If the ITF between any two destinations in a study area is gotten, then a multi-destination network can be built and downstream analysis tasks like tourism demand prediction\cite{kim2022visitor}, community discovery\cite{xu2021characterizing}, trip pattern discovery\cite{hwang2006multicity}, facility suggestion\cite{shih2006network} and multi-attraction network structure analysis~\cite{wang2021spatial,ramadhani2021mapping,hwang2006multicity} can be further conducted. From the perspective of geographic granularity, some of the research are done on inter-city level ITF, for example, Hwang,
Gretzel et al.\cite{hwang2006multicity} examined international tourists’ multi-city trip patterns by using clique detection and statistical analysis on the multi-attraction network created on ITF between American cities. Other researches are done on inter-attraction level ITF, for example, 
Liu, Huang et al.\cite{liu2017application} used network analysis to explore the underlying mechanisms of
tourist attraction network informed by tourist flows between attractions in Xinjiang, China. These two classic researches both used undirected ITF extracted from survey data.

In the past decade, technology developments such as GPS and smart phones have made it possible to get the travel routes of individual tourists, from which the concrete number of ITF can be extracted. However, the GPS trace data of individuals is hard to get due to privacy issues, especially fine-grained data like which attractions the tourists visited. Although there are publicly available data about individual GPS trajectories, such as the GeoLife dataset\cite{zheng2008geolife}, it is hard to be applied to the tourism study field, because the trajectory of tourists is hard to be distinguished from that of other people. In fact, to our knowledge, there is no ITF dataset publicly available. For ITF between attractions, former studies usually extract it from survey data\cite{shih2006network} or social media data like check-in data~\cite{kim2022visitor,wu2014intra} or travel notes data\cite{ramadhani2021mapping}. However, since the social media data can be sparse \cite{lo2011tourism}(i.e., not all cities and places have plenty of social media data to be used, and people may not report all the places they visit on social media), it is common to encounter the problem of missing ITF. Thus, the prediction of ITF is a fundamental problem when the ITF data is not available. This kind of prediction also deepens the understanding of the tourists' mobility patterns. Because the volume of ITF between two attractions is related with factors like distance, ticket price and types of 
attractions, etc..

In this paper, we try to use multi-source accessible information of tourism destinations to predict the inaccessible ITF. Since the ground truth for training the prediction model is inaccessible to us, we use web crawling tools to collect tourism travel notes to the city of Beijing, where the attraction data is relatively sufficient due to large volume of tourist visits to the city. Besides, since tourism functions are non-linear and behave in a dynamical manner, it is difficult to see direct cause and effect between actions\cite{mckercher1999chaos}. We also tried to give a qualitative analysis of whether and how different features like ticket price would influence the volume of ITF.

To predict the ITF, we are inspired by the commuting flow prediction problem~\cite{yao2020spatial,liu2020learning}. Commuting flow prediction deals with predicting realistic flows among locations, given their characteristics and the distance among them, and without any knowledge about the real flows. Analogies can easily be found between the commuting flow prediction problem and the ITF prediction problem, except that the features used in the commuting flow prediction problems are usually population and land-use context instead of information about the tourism attractions, and the objective to be predicted is the volume of commuter flow between regions in a city instead of volume of tourism flow between tourism attractions. Multiple methods have been proposed to deal with the commuting flow prediction problem, including traditional physics-based models like the gravity model; And machine learning-based models such as Deep Gravity model\cite{simini2021deep}, Random Forest\cite{pourebrahim2019trip}, XGBoost\cite{morton2018need}, and Graph Neural Network(GNN) based models, i.e. GMEL\cite{liu2020learning}, SIGCN\cite{yao2020spatial} and RFGCN\cite{yin2022convgcn}. Among these models, the GNN-based models get the state of art performance and can take the implicit features of the interaction network of different regions into consideration. However, when apply the models for commuting flow prediction directly to ITF prediction, the performance is not very well. Thus, we need to revise the models for commuting flow prediction problems to get better performance on ITF prediction problems.

The contribution of the paper is mainly threefold. Firstly, we are the first to use a graph-based deep learning model to predict the ITF between tourism attractions, and our model achieves a Mean Absolute Percentage Error (MAPE) of 0.46 on test data, which outperforms all the existing flow-generating methods applying directly on this problem. Secondly, we explained how the volume of ITF is influenced by different features of attractions by using SHAP (SHapley Additive exPlanations) on a Random Forest model, which is important to understand the nature of tourists' destination choice in the tourism destination system. Thirdly, we created a benchmark dataset of ITF for 246 attractions in Beijing, along with the nine features of these attractions that are used for predicting in our model. This benchmark is publicly available at \href{https://github.com/iq180fq200/tourism_interaction/tree/master/data}{Github}. 

The remainder of the paper is as follows. Section \ref{Literature Review} reviews the relevant works on ITF and the models for the flow prediction problem. Section \ref{definition} presents the problem statement of the ITF prediction task. Section \ref{Research Design} discusses the model we use. The experiments and results of the models and an explanation of the features are then provided in Section \ref{Experimental Results}. Finally, discussions and conclusions are presented in section \ref{Discussions and Future Works} and \ref{Conclusions} respectively.

\section{Literature Review}\label{Literature Review}
\subsection{The Study of Inter-Destination Tourism Flow}
Many tourists go multiple destinations in one trip, and this phenomenon is displayed directly by ITF. A lot of research in tourism management are conducted based on ITF, since it implies things like the relationship between destinations, the tourist behaviour preference, and the structure of multi-destination network. 

From the perspective of directions, the ITF can be devided to directed ITF and undirected ITF. The directed ITF from destination A to destination B refers to the number of tourists that visit B right after visiting A in the same trip; While the undirected ITF between two destinations A and B refers to the number of tourists that visit both A and B in the same trip. 

In previous study, the directed ITF are used to predict tourism flow over time\cite{wang2021multi}, forcast tourism demand\cite{kim2022visitor}, detect tourists' mobility pattern\cite{grinberger2019spatiotemporal}, identify boundary effect\cite{peng2016network} and identify common routes\cite{d2013network}. Another common way of study using directed ITF is to build the multi-destination network whose edges are directed ITF, and then calculate the node structure indicators(i.e. node-centrality indicators and
structural holes indicators) and network structure indicators(i.e. size, network centralization, etc.).The calculated indicators can be used to classify roles of destinations~\cite{d2013network,peng2016network} ,study spatial structure of multi-destination network\cite{gonzalez2015configuration} or compare visitation patterns of tourists with different characteristics(e.g. different origins, purpose, visit duration, e.t.c.)\cite{grinberger2019spatiotemporal}. 

The undirected ITF, on the other hand, is also commonly used for calculating the node structure and network structure indicators, and then doing the same tasks on the basis of calculated indicators as is done on indicators calculated by directed ITF~\cite{wang2021spatial,ramadhani2021mapping,hwang2006multicity}, the only difference is that the multi-destination networks built with undirected ITF are symmetric compared to the networks built with directed ITF. In other word, in multi-destination networks built with undirected ITF, the edge weight from destination i to destination j is the same as that from j to i. Besides calculating the indicators, other methods are also applied to make use of undirected ITF. For example, Xu, Li et al. (2021)\cite{xu2021characterizing} uses a community detection algorithm on the multi-destination network to identify seven destination communities in South Korea; Hwang, Gretzel et al.(2006)\cite{hwang2006multicity} used clique detection to identify strongly-connected destinations in America. In another research by Liu, Huang et al.(2017)\cite{liu2017application}, the Quadratic Assignment Procedure (QAP) was used to test the relationships between region proximity, grade proximity, tenure proximity, and the destination network determined by tourists' free choice movements. Plus, although the undirected ITF has no explicit direction, it can also be used to identify the Directionality of multidestination patterns by using conditional probability comparison\cite{hwang2006multicity}.

From the perspective of granularity, the ITF is divided into two categories, the inter-city level and the inter-attraction level,\footnote{In some articles like \cite{ramadhani2021mapping}, the inter-attraction level is referred to as within-destination level and the inter-city level is referred to as inter-destination level.} In studies of inter-city level ITF~\cite{xu2021characterizing,hwang2006multicity,gonzalez2015configuration}, the whole region containing multiple cities is regarded as a multi-destination system to be studied. While in the studies of inter-attraction level ITF\cite{ramadhani2021mapping,liu2017application,wang2021multi,kim2022visitor,grinberger2019spatiotemporal,shih2006network,peng2016network}, a spatially continuous region(e.g. a city, a province, a river delta) containing multiple tourism attractions is regarded as a system to be studied.

In previous studies, the ITF mainly comes from four data sources, i.e., mobile positioning data, GPS data, survey data, and social media data. The mobile positioning data could capture the location footprints of large populations and is precise and fine-grained. However, such data are difficult to acquire due to privacy issues, and only a few researchers who have access to such data can do research based on it\cite{xu2021characterizing,xu2022understanding}. The GPS data are usually collected by location tracking applications installed on volunteers' phones, which means it can only cover a very limited number of tourists.\cite{grinberger2019spatiotemporal} The survey data is the most widely used way to collect ITF, but conducting surveys is time-consuming and still can only cover a little number of tourists.\cite{gonzalez2015configuration,d2013network,shih2006network,hwang2006multicity} Another commonly used data is social media data. One kind of social media data is Volunteered Geographic Information (VGI) data used in \cite{kim2022visitor,wang2021multi}. It is data published by users and include embedded geographical metadata, like geotagged tweets on Twitter and photos on Flickr. Another kind of social media data used for extracting ITF is tourism notes\cite{ramadhani2021mapping,wang2021spatial,peng2016network}, which, compared to VGI data, are more targeted on tourists. Although being easy to get and covering large crowds, the social media data can be sparse \cite{lo2011tourism},i.e. not all cities and places have plenty of social media data to be used, and people may not report all the places they visit on social media. Thus, predicting missing ITF is important when accurate volume of ITF is not available.

By analyzing ITF, previous studies have found several factors that would influence its volume. Since the factors concerned for inter-city level ITF and inter-attraction level ITF vary a lot, here we mainly discuss the latter, which is the focus of our study. From the perspective of tourists' characteristics, \cite{ramadhani2021mapping} shows that tourists from different original countries result in different ITF patterns; \cite{grinberger2019spatiotemporal} shows that factors like religion, visiting times(i.e., repeated visitor, first visitor), age and purpose of tourists also influence ITF. And \cite{xu2022understanding} finds that the ITF patterns vary in different duration of stay and lengths of tour trajectory of individual tourists. From the perspective of tourism attractions' characteristics, \cite{alvarez2020analysing} finds that there is a visitor flow spillover effect on neighboring attractions, meaning that the distance between attractions would influence the ITF between them; \cite{liu2017application} shows that the proximity of two attractions in their region, level, and time of getting the level would influence the ITF between them; \cite{peng2016network} finds that the provincial-administrative boundary would have a boundary-shielding effect on the ITF between attractions that belong to different provinces; and \cite{do2021determinants} finds that the diversity of multi-attractions and the perceived costs would influence the ITF. However, none of the study integrate all of the characteristics of attractions mentioned above together; and the popularity of attractions was not considered; Besides, although regression models like logistic regression, polynomial regression, and QAP are used to regress the ITF; their main purpose was to explain the influence of independent variables instead of predicting ITF precisely, thus, their model's performance in predicting is limited.

\subsection{Commuting Flow Prediction Problem}
The prediction of ITF is similar to the commuting flow prediction problem, which is to predict realistic commuter flows among locations, given their characteristics and the distance among them, and without any knowledge about the real flows. Traditionally, the gravity\cite{masucci2013gravity} and radiation models\cite{zipf1946p} are the most commonly used ones for this task. The gravity model is based on the assumption that the number of travelers between two locations increases with the locations' population while decreasing with the distance between them; the Radiation model replaces the distance feature in Gravity model with the intervening opportunities, which considers more variables. However, the traditional formula-based models typically have a strict structural form and a limited number of input variables, which limits their ability to predict commuting flows.

With a better ability to capture nonlinear relationships, machine learning models are used for the commuting flow prediction problem. For example, Morton et al.\cite{morton2018need} find that the XGBoost model performs better on this problem compared to traditional models; \cite{pourebrahim2019trip} compare the gravity model, neural networks, and random forest model on Twitter data for predicting commuter flow, and find that the Random Forest model has the best performance. \cite{simini2021deep} put forward a deep gravity model with good interpretability that improved the performance of the traditional gravity model a lot.

Recently, another type of learning model, Graph Neural Network (GNN) based models, has been used for the commuter flow prediction problem. Liu et al. \cite{liu2020learning} put forward a GAT(GNN with Attention) model, GMEL, that generates two embeddings for every node on a geo-adjacency network, and then uses Dismult as the score function. The model effectively captured the spatial correlation
from geographic contextual information. \cite{yao2020spatial} put forward a GCN(Graph Convolutional Netxork) model, SI-GCN, with GCN as encoder and Bilinear as score function, which reaches a good performance on the T-Drive dataset. Yin et al.\cite{yin2022convgcn} put forward a GCN model with a multilayer perceptron as the score function and a random forest as a refinement layer on the task, which effectively improved the prediction accuracy. Most recently, Yin et al. \cite{yin2022convgcn} put forward a GCN model, ConvGCN-RF, with multilayer perceptron as a score function and a random forest as a refinement layer on the task, which effectively improved the prediction accuracy. 

However, applying the above models directly on the ITF prediction problem has some drawbacks. For non-GNN-based models like deep gravity or Random Forest, they cannot capture the interaction relationships of multiple(i.e., more than two) attractions as a system. For example, if we already know that two attractions, $A_i$ and $A_j$, has a strong interaction, which means many people choose to visit both of them in one trip; and another attraction, $A_k$, is on the way from $A_i$ to $A_j$. Then we may guess many visitors would also visit $A_k$ as a ``drop by visit" when they go from $A_i$ to $A_j$. Thus, the volume of ITF between $A_i$ and $A_k$ may also be large. As for the GNN-based models, the GMEL must be applied to continuous space, while the tourism attractions are discretely scattered areas in the city. The SI-GCN model can use the coordinates as input features to take the relative positions of attractions into consideration, but this method could not take consideration of the distance feature explicitly. While the RFGCN, which uses an MLP as the score function, could not explicitly consider the corresponding features of $A_i$ and $A_j$ jointly when predicting the ITF between them.

\section{Preliminaries and Problem Definition}\label{definition}
In this study, we mainly study \textbf{the prediction problem of undirected ITF at the inter-attraction level }, although the proposed model can also be extended to predicting directed ITF, as is discussed in Section \ref{Discussions and Future Works}. In this section we first introduce preliminary definitions used in this research and then define the problem of our research.\\

\textbf{Attraction \textbf{A}:} An attraction is a tourism place with an independent entrance and a clearly-defined geographic area. For example, the Forbidden City is an attraction while the Taihe Palace is not, because it is within the Forbidden City and has no independent entrance. Each attraction has the features of locations, ticket price, area, type etc., which can be represented as a feature vector. \\

\textbf{Trip T:} A trip is defined as a sorted series of attractions that a tourist visit in one journey. Here we regard the time span of a journey as from arriving to a city to leaving a city.\\

\textbf{Inter-attraction undirected ITF: } Given the trips in a certain time period (like recent ten years), an inter-attraction undirected ITF between two attractions $A_i$ and $A_j$ is the total number of trips that tourists have visited both $A_i$ and $A_j$ regardless of their visit order. In the rest of the paper, we use ITF instead of inter-attraction undirected ITF for simplicity.\\

\textbf{Interaction Graph G(A,E):} The interaction graph is an unweighted undirected graph with nodes representing attractions while edges E representing the status of strong connections (0 or 1) between attractions. In our experiments, we assume that we have already known some of the ITF while the others are missing, so we determine if there exists a strong connection between two attractions by judging if there is a known ITF that is larger than a threshold between them. However, in practice, we can still generate an interaction graph even if no ITF is available at all. One of the possible ways is to use travel-agency-recommended-routes, as is done in \cite{peng2016network}, then the attractions in the same recommended route would have edges connecting them in the Interaction Graph.\\

\textbf{Problem Statement:}  When given an Interaction Graph G = (V,E), learn a model M that is effective to accurately predict the ITF between any two attractions i and j ($ITF_{ij} = M(G,i,j)$), with minimal percentage deviation from the true value.

\section{Methodology}\label{Research Design}
\subsection{Preprocessing}
First, we build an interaction Graph G. We get the features (like location, area, ticket price, etc.) of each tourism attraction. For categorical features, we use ordinal encoding to convert them to numerical forms. Then we normalize all the features into the scale of 0 to 1 and concatenate them to get the feature vector of every attraction. The edges for G is gotten by measures mentioned in the definition of Interaction Graph in section \ref{definition}.

\subsection{Model}
\begin{figure}[ht]
 
\centering
\includegraphics[scale=0.5]{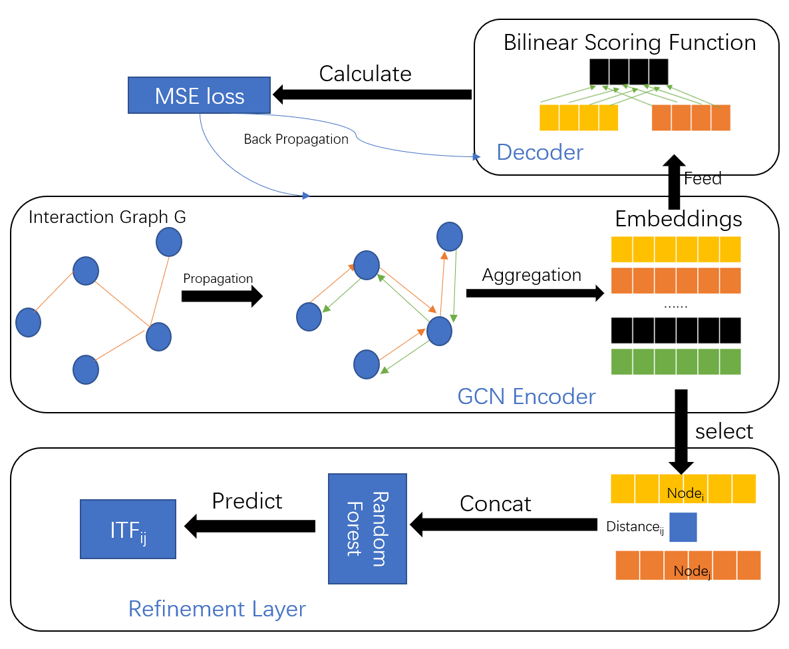}
\caption{Structure of the proposed SI-GCN-RF model}
\label{model}
 
\end{figure}

The model we use in this research is similar to the Spatial Interaction Graph Convolutional Network (SI-GCN) model proposed by \cite{yao2020spatial}; the difference is that we add an extra Random Forest Layer as a refinement layer on top of the SI-GCN architecture so that the distance feature can be considered explicitly; Thus, we call our model SI-GCN-RF(Spatial Interaction Graph Convolutional Network with Random Forest refinement)
As Figure \ref{model} shows, the SI-GCN-RF model consists of three parts, the Encoder, the Decoder, and the Refinement layer.

\subsubsection{Encoder}
The encoder is a graph embedding with $L$ Graph Convolution layers\cite{schlichtkrull2018modeling}.  Every layer in the encoder contains two steps: For each node in $G$, the message passing step collects neighboring nodes' embedding vector(i.e. state), and the embedding updating step calculates the new embedding vector of the node itself. To combine the two steps together, we get the mathematical definition of a Graph Convolution layer of our model as follow:
\[h_v^{l+1} = \sigma (W_l\sum_{u\in N(v)}\frac{h_u^{(l)}}{|N(v)|}+B_lh_v^{(l)} ),(\forall l \in {0,...,L-1})\]
where $h_v^{(l)}$is the embedding for node $v$ in the $l^{th}$ layer, and $h_v^{(0)}$ is the original feature vector of node $v$ in G. $N(v)$ is all the nodes connected with node $v$, $|N(v)|$ is the cardinality of $N(v)$. Function $\sigma$ is the relu non-linearity function, while $W_l$ and $B_l$ are both learnable parameters.

\subsubsection{Decoder}
When dealing with edge prediction by GCN, the Decoder is a scoring function that map the embeddings of two nodes into a real number $R$. In our case, the number $R$ represents the predicted ITF between the two attractions represented by the two nodes. Here, we use Bilinear scoring function\cite{yang2014embedding}, which is commonly used for GCN. In our research, its function is as follow:
\begin{equation}
f(u,v) = \sum_i^{} E_i^u R_i E_i^v 
\end{equation}
where the $f(u,v)$ is the predicted ITF between node u and node v; $E_i^u$ is the scalar value on the $i^{th}$ position of the embedding of node u. R is a learnable vector whose length is the same as the node embedding generated by the encoder and $R_i$ is the element of the $i^{th}$ position of R. 

The Bilinear scoring functions has two advantages over other scoring functions like Distance\cite{bordes2011learning}, Single Layer\cite{socher2013reasoning} or Multilayer Perceptron (MLP). First, it is symmetric. That is to say, the predicted ITF is the same from $A_i$ to $A_j$ as from $A_j$ to $A_i$ due to commutative law of multiplication. And the (undirected inter-attraction level) ITF is also symmetric according to its definition. Second, it can consider the corresponding features of $A_j$ and $A_i$  jointly when predicting the ITF between them. For example, when deciding which attractions to visit on the same trip, tourists may consider visiting attractions of the same type because they have a fixed taste; or they may estimate the ticket price of multiple attractions at the same time to limit their total cost for the trip. To consider this kind of relation, we need the model to identify and consider the corresponding features of the two attractions jointly when predicting the ITF between them. And the Bilinear scoring function can do it, because the numbers that are multiplied in the same item are from the same position of the embedding of $A_i$ and $A_j$.

The output of the decoder is then used to train the parameters in the encoder and the decoder using back propagation, and we use MSE as the loss function:
\[MSE = \frac{1}{n}\sum_{i,j}(y^r_{ij}-y^p_{ij})^2\]
where the real value and predicted value of ITF between two attractions $A_i$ and $A_j$ are represented by $y^r_{ij}$ and $y^p_{ij}$ respectively; and n represents the total number of ITF predicted.

We should note that the output of the decoder is only used to train the model. However, the final output prediction is made by the refinement layer as explained followed.

\subsubsection{Refinement Layer}
After learning the embedding vectors of nodes, we train a refinement layer to improve the performance. The refinement layer is a Random Forest. To get the ITF between $A_i$ and $A_j$, we concatenate their learned embeddings generated by the Encoder together with the distance between them, and thus get a long vector. We then feed the long vector into the Random Forest model to get the predicted ITF. Compared to using the output of the decoder directly as the final prediction, using the refinement layer to concatenate the distance with generated node embeddings can explicitly make use of the distance information, because the distance is related with two attractions and thus can not be represented by the feature vector for individual attractions.

\subsection{Explain the Features}
Since the GNN-based models incorporate both graph structure and feature information, they lead to complex models and predictions made by GNNs is hard to explain~\cite{yuan2022explainability}. However, non-GNN-based models like random forests can be easily explained by existing explainable AI frameworks such as SHAP. So to account for how different features influence the volume of ITF, we first train a Random Forest(RF) model to predict ITF from existing features, than we use SHAP to explain the trained RF model.

SHAP applies a game theoretic approach to explain the output of machine learning models. SHAP values are used to determine feature importance and whether it influences the predicted result negatively or positively. The interpretation of the SHAP value $\phi_j$ for variable value j is: the value of the $jth$ variable contributed to the prediction of a particular instance compared to the average prediction for the dataset. For example, if we find that for variable "distance", when its value is large, its SHAP value is negative; and when its value is small, its SHAP value is positive. We can determine that the model has learned that the distance feature has a negative correlation with the predicted ITF.

\section{Experiments}\label{Experimental Results}
\subsection{Experimental Settings}
\subsubsection{Datasets}
The study uses multiple sources of datasets to get both the training and testing ITF dataset and the features of attractions. 

\begin{figure}[h]
\begin{minipage}[b]
{0.45\linewidth}
\centering
\includegraphics[scale=2.0]{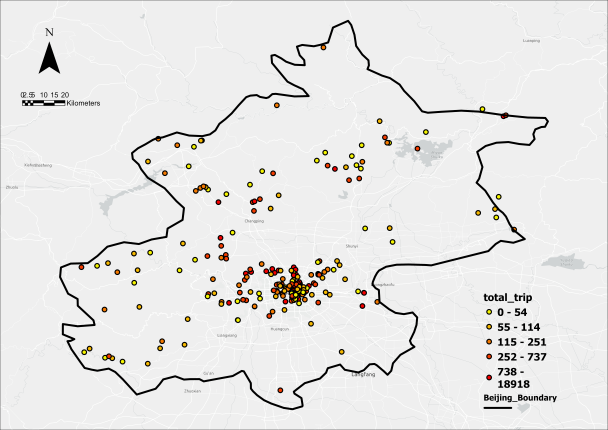}
\caption{The spatial distribution of tourism attractions in this study}
\label{Attractions}
\end{minipage}
\begin{minipage}[b]{0.49\linewidth}
\centering
\includegraphics[scale=0.7]{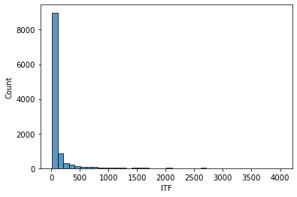}
\caption{The histogram of ITF frequency distribution}
\label{data distribution}
\end{minipage}
\end{figure}

The first is the ITF dataset, which is used as training and testing data for our experiments. Note that for the strict definition of ITF, we need to collect the tourists' trips of all the tourists visiting a region (such as a city) in a certain time period, and then extract the ITF from the whole set of trips. However, such a fine-grained data is not available publicly. Thus, in practice, we use the following procedure to get the ITF: First, we downloaded 68594 travel diary notes to the city of Beijing, China from the Mafengwo \footnote{http://www.mafengwo.cn/
}, which is the largest website for sharing travel diaries from tourists in China. The timespan of the data is from September 2012 to July 2021. Second, we collected the place names of 300 main attractions in Beijing from the Mafengwo. After manual data clean to filter out places that don't align with the definition of attraction in section \ref{definition}, we kept 246 attractions in total for further research, the spatial distribution of which is shown in figure \ref{Attractions}. Third, we created a text-matching template for the attractions' names, which considered the abbreviation and alias of every name. Next, we used the text-matching template to extract the trip of individual tourists from their travel notes. In case some tourists may post multiple travel notes for a single trip, we merged such travel notes posted by the same tourist within five days before extracting the trip. And trips with visited attractions less than two are dismissed. We got 30860 trips in total with the mean number of attractions visited in one trip as 7.34 and the median as 5.0. Finally, we calculated the ITF between any two attractions from the trips we got. The numerical distribution of ITF is shown in figure \ref{data distribution}, with a mean of 172, and a standard deviation of 603.

The second dataset is the features/attributes of tourism attractions, which is used as independent variables in our predictive model. There are nine features in total that we took into consideration for estimating the quality, coordinates, popularity, and other characteristics of attractions, which is shown in Table \ref{features}. While other features were gotten directly from the data sources, the areas of attractions are gotten by first downloading their boundary coordinates from the Amap\footnote{https://www.amap.com/
}, then calculating their areas according to their boundaries with GeoPandas.

\begin{table}[h]
\centering
\caption{The data sources of tourism attraction features}
\begin{tabular}{cccc} \hline
feature& data source& intention& description\\
\hline
lon& Amap& position& center longitude of attraction\\
lat& Amap& position& center latitude of attraction\\
area& Amap& characteristic& area of attraction\\
adname& Amap& position& administration district\\
ticket\_price& Ctrip& characteristic& ticket price of attraction\\
type& Ctrip& characteristic& type of attractionb\\
ranking& Ctrip& quality& average ranking\\
Comment Number& Ctrip& popularity& comment number of attraction\\
Level& MBCT& quality& official level estimation of attraction\\
Estimated Visiting Time& Qunar& characteristic& estimated average visiting time\\
\hline
\end{tabular}
\label{features} 
\end{table}

Among these features, the "Adname" is the administration district that the attraction belongs to. The "comment number" and "ranking" of attractions are collected from the Ctrip\footnote{https://ctrip.com/
}, which is among the largest public ranking website for attractions in China. The "comment number" is an important feature to show the popularity of an attraction, and the "ranking" means the average ranking of an attraction, which is used to evaluate the overall quality of an attraction. Note that the "ranking" is more reliable when the "comment number" is getting higher, which means more people have ranked the attraction. The evaluation system of the "level" feature is an official system issued by the China National Tourism Administration to evaluate the value and quality of attractions. In this system, all attractions in China are evaluated to be either no-level, 2A, 3A, 4A, or 5A, with a larger number meaning better quality. In our research, we downloaded the "level" of all the attractions we studied from the website of Beijing Municipal Bureau of Culture and Tourism\footnote{http://whlyj.beijing.gov.cn/
}. We assigned the "level" feature of 2A to 5A attractions with 2 to 5 respectively; and assign the "level" feature of no-level attractions with 1. The "type" feature of the attractions is a categorical feature includes six categories: historical sites, natural scenery, zoos \& arboretums, amusement park,city sightseeing, and exhibition center \& museum. The "type" information is also gotten from the Ctrip website. The "mean visit time" is the mean time spent to visit a certain attraction evaluated by the Qunar \footnote{https://travel.qunar.com/}, another large tourism website in China (similar to Expedia/Booking websites in the US). When training the model, we divide the total ITF set to 60\%, 20\%, 20\% as the training sets, validation sets, and testing sets respectively.

\subsubsection{Baselines}
The ITF prediction problem is similar to commuter flow generating problems, so the methods used for commuter flow generation problem can be directly applied to it. Among these, We used representative non-GNN based learning methods (i.e., Random Forest and Deep Gravity) and the state-of-the-art GNN-based learning methods (i.e., SI-GCN and GCN-RF ) as baselines to compare.\\

\textbf{Random Forest.} To get the ITF between two Attractions $A_i$ and $A_j$, we concatenate their features and the distance between them. For catagorical features, we use ordinal encoding to convert them to numerical forms. We then feed the concatenated feature vector into the Random Forest model to get the predicted ITF between $A_i$ and $A_j$.\\

\textbf{Deep Gravity.} After we get the input vector as the same in Random Forest, we feed it into a feed-forward neural network with 15 hidden layers with LeakyReLu activation functions. The output of the last hidden layer is the predicted ITF.\\

\textbf{SI-GCN.} The SI-GCN model is a SI-GCN-RF model without the random forest based refinement layer. That is to say, the output of the Bilinear scoring function is the predicted ITF of the model.\\

\textbf{GCN-RF.} The GCN-RF model uses a standard GCN as the encoder and an MLP as the decoder to train the GCN model. Then it uses the embeddings generated by the trained GCN to train a random forest model as refinement layer.\\

\textbf{SI-GCN-RF(no edge).} To determine the effectiveness of the interaction graph used in SI-GCN-RF, we also test a model whose Graph has no edge, and we call it \textbf{SI-GCN-RF(no edge)}. The components of this model are the same as SI-GCN-RF, except that there is no message passing between neighboring nodes when training the encoder. 
\begin{figure}[ht]
\centering
\includegraphics[scale=0.5]{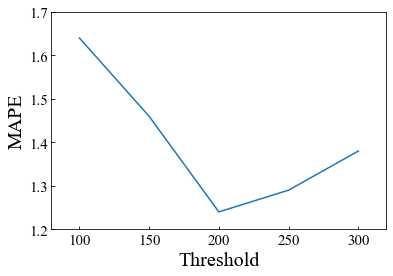}
\caption{The sensitivity analysis result of the threshold of the interaction graph}
\label{Sensitivity Analysis Result}
\end{figure}
\subsubsection{Hyperparameters}
We set the hyperparameters based on
the model performance. Specifically, we set the ITF threshold of the interaction graph as 200, for the encoder-decoder part can have a best performance on this threshold as is shown in figure \ref{Sensitivity Analysis Result}. The dim $D$ of the linear layer in preprocessing part and the dim of $B_l$ in the encoder are both 500, while $W_l$ in the encoder is a 500*500 matrix. We adopt Adam optimizer for model training. In addition, the batch size is set to full batch, the epoch is set to 50000, with an early stopping mechanism whose patience is 500 epochs. The max norm of the gradient clip is set to 1.0, the initial learning rate is set to 0.02, and the dropout is set to 0.0 (i.e., no drop out). The number of estimator for the random forest refinement layer is 30, while the max depth is 25. All the hyperparameters of baseline models are also tuned to have the best performance. We implement and compare all the models with Sklearn and Pytorch.
\subsubsection{Evaluation metrics}
We adopt three metrics to evaluate the performance of the models: Mean Square Error (MSE), Mean Absolute Percentage Error (MAPE) and Common Part of Commuters (CPC). MSE is to evaluate the absolute deviation from the true value to the predicted value in regression problems; MAPE is is a metric to evaluate the relative difference between prediction and ground truth; and CPC is the most commonly used metric to evaluate the performance of flow generation models. For MSE and MAPE, the smaller value means the better performance; while for CPC, the larger means the better. While MSE has no range constraint; MAPE and CPC can only have values between 0 to 1. The mathematical definition of the MAPE and CPC are as follow:
\[MAPE = \frac{1}{n}\sum_{i,j}|\frac{y^r_{ij}-y^p_{ij}}{y^r_{ij}}|\]
\[CPC = \frac{2\Sigma_{i,j}min(y^r_{ij},y^p_{ij})}{\Sigma_{i,j}y^r_{ij} + \Sigma_{i,j}y^p_{ij}}\]

where the real value and predicted value of ITF are represented by $y^r$ and $y^p$ respectively; and n represents the total number of ITF predicted.

\begin{table}[h]
\centering
\caption{The model comparison results}
\begin{tabular}{ccccccc} \hline
Model& MSE& MAPE& CPC& GNN Based& Scoring Function& Refinement\\
\hline
Random Forest& 103780.59& 1.99& 0.68& no& /& /\\
Deep Gravity& 363613.97& 2.71& 0.35& no& /& /\\
GCN-RF& 499576.74& 5.68& 0.31& yes& MLP& yes\\
SI-GCN& 11665.91& 1.25& 0.85& yes& Bilinear& no\\
\textbf{SIGCN-RF}& 21420.55& 0.46& 0.89& yes& Bilinear& yes\\
\textbf{SIGCN-RF(no edge)}& 22335.29& 0.67& 0.88& yes(no message passing)& Bilinear& no\\
\hline
\end{tabular}
\label{performance} 
\end{table}

\begin{figure}[h]
\begin{minipage}[b]{0.5\linewidth}
\centering
\includegraphics[scale=3]{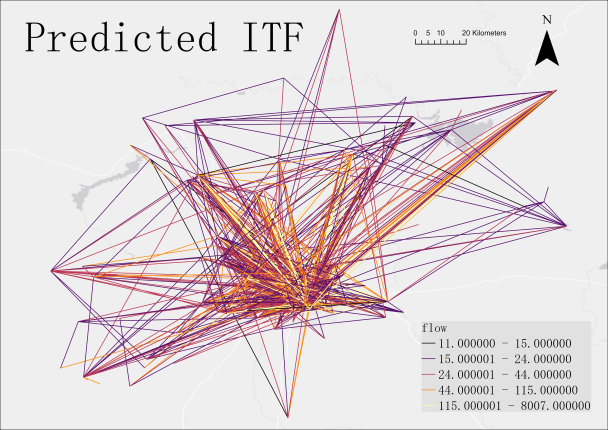}
\end{minipage}
\begin{minipage}[b]{0.5\linewidth}
\centering
\includegraphics[scale=3]{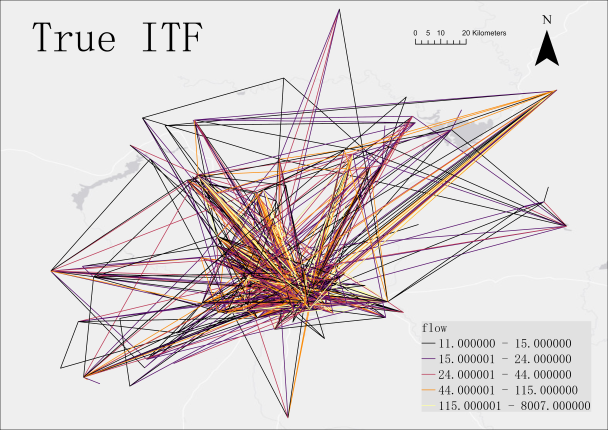}
\end{minipage}
\caption{The visualized results of  predicted ITF compared to the true ITF values}
\label{fig:flowmap}
\end{figure}

\subsection{Performance Evaluation}
We compare the proposed model SI-GCN-RF and all baselines in terms of the prediction performance (i.e., MAPE, MSE and CPC). Table \ref{performance} reports
the model comparison results. Since the MSE is an absolute metrics and thus has a strong bias towards instance with larger true value, it is least representative when evaluating the model's performance. So we mainly use MAPE and CPC when analyzing the performance in the following parts. From the results, we observed that the models with a Bilinear scoring function after node embedding have better performance with CPC all larger than 0.85 and MAPE all less than 1.3, compared with other models whose CPC are all less than 0.7 and MAPE larger than 1.95. That is because, the Bilinear scoring function for node embeddings successfully extracts the collaborate influence of corresponding features and also maintains the symmetric feature of ITF. Among these, the SIGCN-RF with both massage passing in Interaction graph, self-updating, Bilinear scoring function and Random Forest Refinement layer has the best performance, with an MAPE of 0.46 and CPC of 0.89. The performance of SIGCN-RF(no edge) is worse than SIGCN-RF with MAPE larger by 0.2, which means the interaction graph successfully captures the interaction intensity features of multiple attractions, thus increases the prediction accuracy. When comparing the SI-GCN with SIGCN-RF, we see that the Random Forest Layer successfully improves the performance of the model by decreasing the MAPE by over 0.7. That is because, the random forest layer has the distance feature input explicitly, while the SI-GCN can only consider the distance by implicitly using the coordinate features as node vectors. For non-GNN based models, the random forest model has a significantly better performance than the deep gravity with CPC increased by over 0.3, which is also one of the reasons of why we choose RF for explaining features. Another interesting thing is that the GCN-RF model with MLP as its scoring function has the worst performance, which may because that the MLP is not suitable for predicting undirected ITF with volume from $A_i$ to $A_j$ equal to that from $A_j$ to $A_i$. This phenomenon may also partly explain the limited performance of the deep gravity model, whose backbone is also MLP.
\begin{figure}[h]
\begin{minipage}{0.5\linewidth}
    \centering
\includegraphics[scale=0.6]{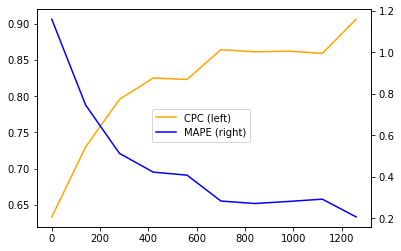}
\caption{The prediction accuracy over different true values}
\label{range_accuracy}
\end{minipage}
\begin{minipage}{0.5\linewidth}
    \centering
\includegraphics[scale=0.35]{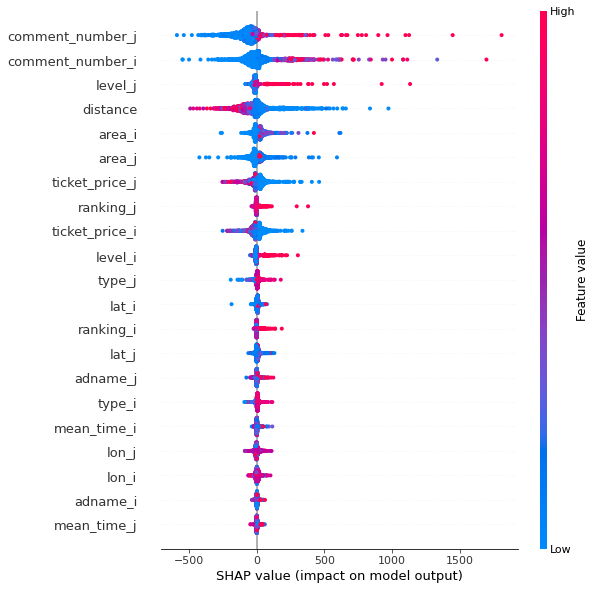}
\caption{The global result of SHAP analysis}
\label{rf_global_shap}
\end{minipage}
\end{figure}

A visualized result of the predicted ITF compare with the true ITF is presented in figure \ref{fig:flowmap} with quantile classified symbology. Besides, we notice that the relative prediction accuracy on larger true ITF values are better, as is shown in figure \ref{range_accuracy}.

\subsection{Feature Explaination}\label{feature explaination}
When using SHAP to explain the influence of different attractions' features on ITF in the RF model, we get the SHAP values for every attraction feature in table \ref{features}, in every prediction iteration. The result is shown in figure \ref{rf_global_shap}, with the features sorted by their maximum SHAP values.

\begin{figure}[h]
\begin{minipage}[b]{0.5\linewidth}
\centering
\includegraphics[scale=0.6]{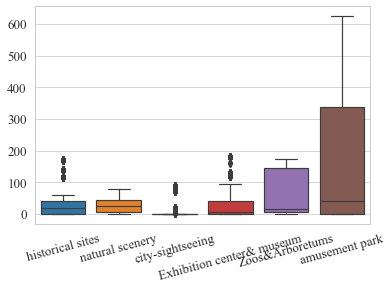}
\caption{The ticket price over different attraction types}
\label{price over types}
\end{minipage}
\begin{minipage}[b]{0.5\linewidth}
\centering
\includegraphics[scale=0.6]{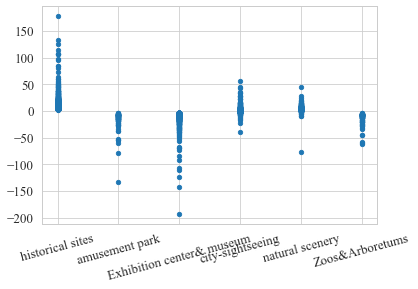}
\caption{The SHAP value over feature $type$}
\label{SHAP_type}
\end{minipage}
\end{figure}

From the result shown in figure \ref{rf_global_shap}, we have four observations. First, popularity, quality, and distance are the main influential factors in the predicted volume of ITF for the RF model. Second, the comment number of attractions, which represents the popularity of attractions, and the level and average ranking, which represent the quality of attractions, have a explicit positive relation with ITF. This is because people tend to go to attractions that are popular and have a good reputation. Third, the distance of attractions has a negative relation with ITF, the nearer the two attractions are, the larger the ITF between them is. This verified that there is a spill-over effect of neighboring tourism attractions. Fourth, high ticket price usually leads to low ITF. That may be because, the attractions with a high ticket price are usually amusement parks or Zoos \& Arboretums as is shown in figure \ref{price over types}, compared to free attractions, fewer visitors may be willing to pay a high price to those places. Also, compared to historical sites or city sightseeing spots, these attractions show fewer unique characteristics of the city, thus are less possible to attract visitors.

\begin{figure}[H]
\begin{minipage}[b]{0.5\linewidth}
\includegraphics[scale=0.6]{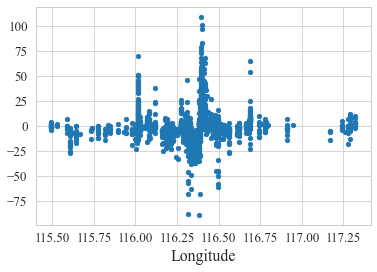}
\end{minipage}
\begin{minipage}[b]{0.5\linewidth}
\includegraphics[scale=0.6]{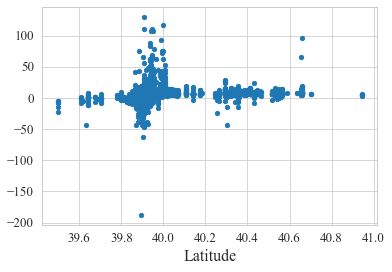}
\end{minipage}
\caption{The SHAP value over feature $coordinates$}
\label{coordinate-SHAP}
\end{figure}  
\begin{figure}[]
\begin{minipage}[b]{0.5\linewidth}
\centering
\includegraphics[scale=0.6]{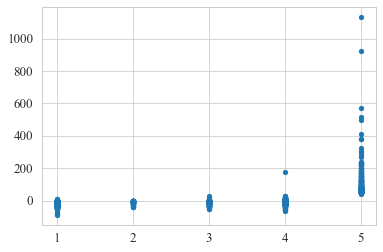}
\caption{The SHAP value over feature $level$}
\label{SHAP_level}
\end{minipage}
\begin{minipage}[b]{0.5\linewidth}
\centering
\includegraphics[scale=2.5]{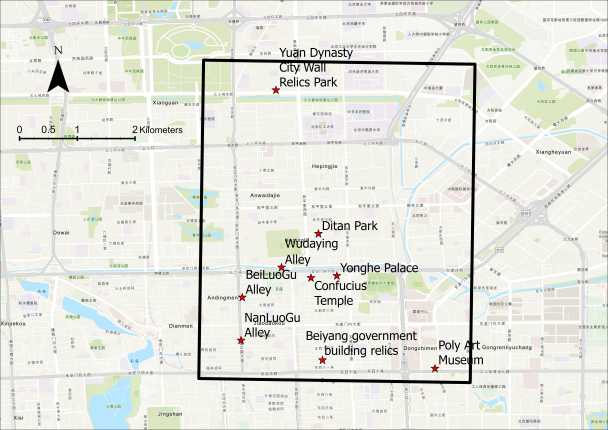}
\caption{The geographic area for high SHAP values}
\label{area}
\end{minipage}
\end{figure}

We also examine the SHAP for features that shows no implicit influential tendencies in figure \ref{rf_global_shap} (i.e. type, coordinates, area, adname, mean\_time)  in detail. Note that in the RF model, when predicting the ITF between two attractions, their feature vectors are concatenated and then fed into the model, thus, a feature is represented by both the feature of $Attraction_i$ and $Attraction_j$, (like the $ticket\_price$ feature is represented by both $ticket\_price_i$ and $ticket\_price_j$), we include the SHAP value of the same feature for both $Attraction_i$ and $Attraction_j$ when analyzing a particular feature like ticket\_price. When we examine the SHAP value for feature "type" in figure \ref{SHAP_type}, we can find that historical sites have an explicit positive influence on ITF, whereas amusement parks, exhibition center \& museum has an explicit negative influence. 
When observing the SHAP value for coordinate features as is shown in figure \ref{coordinate-SHAP}, we find an interesting phenomenon. When the longitude of attraction is between 116.396E and 116.438E; and the latitude is between 39.93N to 39.98N, the coordinate will have a positive effect on the ITF. If we map these areas on the map, as is shown in figure \ref{area}, we can see that this area is the NorthEastern area of downtown Beijing, with a lot of neighboring Hutongs (narrow lanes or alleyways in a traditional residential area of a Chinese city) there. 

Another interesting phenomenon is that if we examine the SHAP value for feature "level", we can see from figure \ref{SHAP_level} that only the 5A attractions will have an explicit positive influence on ITF. In China, the 5A attractions are rare attractions with high value, and they are ranked by experts strictly. Their high quality is contributing to a large ITF that travels to them.

However, the adname, estimated visiting time, and area of attractions have no explicit tendencies found.

\section{Discussions}\label{Discussions and Future Works}
\subsection{Extend the model to directed ITF}
Our proposed SI-GCN-RF model can be modified to predict the directed ITF. To achieve this goal, the structure of the encoder can be the same, with the Interaction Graph G as a directed binary graph, and the  $N(v)$ in the encoder meaning the neighboring vectors who have an edge connected to the node $v$. In the problem of predicting directed ITF, the scoring function in the decoder can be modified to be a Dismult:
\[f(u,v) = E^u R E^v\]
where the $f(u,v)$ is the predicted ITF from node u to node v; $E^u$ and $E^v$ are the embedding vector of node u and v respectively, and $R$ is a learnable square matrix. Since $R$ is not necessarily a symmetric matrix, the Dismult decoder can allow the flow from node $u$ to node $v$ different than that from node $v$ to node $u$. We have not tested this extended model due to the data limitation. Tourism notes are not necessarily written in the order of visiting, thus are more suitable for undirected ITF. The test of the extended model or other better models on directional ITF prediction is left to future works.

\subsection{Explain the influence of Interaction Graph}
We should note that SHAP can only explain explicit features like the distance between $A_i$ and $A_j$ and the features defined on individual attractions like the ticket price. However, it cannot consider the implicit features of the structure of the interaction graph nor the joint effect of corresponding features. 

Recently, several explainable AI methods on GNN have been put forward\cite{yuan2022explainability}, and they may be able to explain the prediction models of ITF with GNN involved. We also left it to be a future research focus.


\section{Conclusions and Implications for Tourism Management}\label{Conclusions}
The ITF is a fundamental dataset when modeling the multi-destination tourism system, but it is hard to get due to limitations of data collection techniques (such as the sample size of surveys or the sparsity of social media data) and privacy issues. We put forward a model to predict the inter-attraction-level undirected ITF when the data is not available. The model uses a GCN over the interaction graph to consider the interaction of more than two attractions to capture the implicit influential factors for the volume of ITF like the tend of ``drop by visit", and to make use of nine features of tourism attractions as node vectors to consider the quality, popularity, position, and characteristics like attraction type of every attraction. It also uses the Bilinear as its scoring function to consider the corresponding features of two attractions jointly when predicting the ITF between them. Besides, it uses an RF refinement layer to consider the distance feature between each pair of attractions. The model outperforms all the commuter flow prediction models using directly on the ITF prediction problem and gets a MAPE of 0.46 and a CPC of 0.89. We also discussed how the model can be modified to predict directed ITF. All the data we used in this research is created as a public benchmark.

The prediction model can be used for many downstream tasks that need to understand a multi-destination tourism system by analyzing ITF data, when the dataset is incomplete or unavailable at all. These tasks help improve the efficiency of tourism management. For example, the community detection based on predicted ITF data can be used to design better collaborative marketing strategy; the statistical analysis on the predicted ITF data may help to suggest facility configuration and the design of tour bus within city; downstream regression models used on the predicted ITF can be used to forecast the travel demand of certain tourism destination.

To understand the relationship between different features of attractions and the ITF, we also apply a SHAP analysis framework on an RF white-box model. The result shows that popularity, distance, and quality are the top three important features that influence the predicted volume of ITF in an RF model. Other features like type, price, and coordinates would also exert an influence in different ways. This gives us better understanding of how tourists' visiting choice is influenced by various factors. Main conclusions include but not limited to: Generally, most tourists would choose to visit multiple popular and high-quality destinations in one trip, even though the distance between them may be far; the nearer distance between two attractions will lead to more co-visitation; the influence of 5A level estimation is explicitly greater than 4A or lower level estimation for tourism attractions; the amusement park, exhibition centers with high ticket price generally has less ITF between them and other attractions. These understandings give us suggestions in tourism management, like considering the advertisement of attractions on the way between two popular, high-quality tourism attractions and improve the tourists' impression of ``an unpopular attraction has some connection with a nearby popular 5A attraction". 

Future work will focus on predicting the directed ITF and explaining the influence of the interaction of multiple attractions on the ITF.

\bibliographystyle{plain}
\bibliography{main}

\begin{thebibliography}{10}

\bibitem{alvarez2020analysing}
Marcos Alvarez-Diaz, Beatrice D'Hombres, Claudia Ghisetti, and Nicola
  Pontarollo.
\newblock Analysing domestic tourism flows at the provincial level in spain by
  using spatial gravity models.
\newblock {\em International Journal of Tourism Research}, 22(4):403--415,
  2020.

\bibitem{bordes2011learning}
Antoine Bordes, Jason Weston, Ronan Collobert, and Yoshua Bengio.
\newblock Learning structured embeddings of knowledge bases.
\newblock In {\em Twenty-fifth AAAI conference on artificial intelligence},
  2011.

\bibitem{do2021determinants}
Hao Thi~Kim Do, Dung~Phuong Hoang, and Thuy~Thu Pham.
\newblock Determinants of multi-destination travel in vietnam: a rational
  choice perspective.
\newblock {\em International Journal of Tourism Cities}, 8(2):289--310, 2021.

\bibitem{d2013network}
Rosario D’Agata, Simona Gozzo, and Venera Tomaselli.
\newblock Network analysis approach to map tourism mobility.
\newblock {\em Quality \& quantity}, 47(6):3167--3184, 2013.

\bibitem{gonzalez2015configuration}
Belen Gonzalez-Diaz, Mar Gomez, and Arturo Molina.
\newblock Configuration of the hotel and non-hotel accommodations: An empirical
  approach using network analysis.
\newblock {\em International Journal of Hospitality Management}, 48:39--51,
  2015.

\bibitem{grinberger2019spatiotemporal}
A~Yair Grinberger and Noam Shoval.
\newblock Spatiotemporal contingencies in tourists’ intradiurnal mobility
  patterns.
\newblock {\em Journal of Travel Research}, 58(3):512--530, 2019.

\bibitem{hwang2006multicity}
Yeong-Hyeon Hwang, Ulrike Gretzel, and Daniel~R Fesenmaier.
\newblock Multicity trip patterns: Tourists to the united states.
\newblock {\em Annals of Tourism Research}, 33(4):1057--1078, 2006.

\bibitem{kim2022visitor}
Yoo~Ri Kim, Anyu Liu, Jason Stienmetz, and Yining Chen.
\newblock Visitor flow spillover effects on attraction demand: A spatial
  econometric model with multisource data.
\newblock {\em Tourism Management}, 88:104432, 2022.

\bibitem{liu2017application}
Bing Liu, Songshan~Sam Huang, and Hui Fu.
\newblock An application of network analysis on tourist attractions: The case
  of xinjiang, china.
\newblock {\em Tourism Management}, 58:132--141, 2017.

\bibitem{liu2020learning}
Zhicheng Liu, Fabio Miranda, Weiting Xiong, Junyan Yang, Qiao Wang, and Claudio
  Silva.
\newblock Learning geo-contextual embeddings for commuting flow prediction.
\newblock In {\em Proceedings of the AAAI conference on artificial
  intelligence}, volume~34, pages 808--816, 2020.

\bibitem{lo2011tourism}
Iris~Sheungting Lo, Bob McKercher, Ada Lo, Catherine Cheung, and Rob Law.
\newblock Tourism and online photography.
\newblock {\em Tourism management}, 32(4):725--731, 2011.

\bibitem{masucci2013gravity}
A~Paolo Masucci, Joan Serras, Anders Johansson, and Michael Batty.
\newblock Gravity versus radiation models: On the importance of scale and
  heterogeneity in commuting flows.
\newblock {\em Physical Review E}, 88(2):022812, 2013.

\bibitem{mckercher1999chaos}
Bob McKercher.
\newblock A chaos approach to tourism.
\newblock {\em Tourism management}, 20(4):425--434, 1999.

\bibitem{morton2018need}
April Morton, Jesse Piburn, and Nicholas Nagle.
\newblock Need a boost? a comparison of traditional commuting models with the
  xgboost model for predicting commuting flows (short paper).
\newblock In {\em 10th International Conference on Geographic Information
  Science (GIScience 2018)}. Schloss Dagstuhl-Leibniz-Zentrum fuer Informatik,
  2018.

\bibitem{peng2016network}
Hongsong Peng, Jinhe Zhang, Zehua Liu, Lin Lu, and Lu~Yang.
\newblock Network analysis of tourist flows: a cross-provincial boundary
  perspective.
\newblock {\em Tourism Geographies}, 18(5):561--586, 2016.

\bibitem{pourebrahim2019trip}
Nastaran Pourebrahim, Selima Sultana, Amirreza Niakanlahiji, and Jean-Claude
  Thill.
\newblock Trip distribution modeling with twitter data.
\newblock {\em Computers, Environment and Urban Systems}, 77:101354, 2019.

\bibitem{ramadhani2021mapping}
Dian~Puteri Ramadhani, Andry Alamsyah, Muhammad~Nashir Atmaja, and Joe
  Nathan~CG Panjaitan.
\newblock Mapping complex tourist destination preferences: Network
  perspectives.
\newblock In {\em 2021 9th International Conference on Information and
  Communication Technology (ICoICT)}, pages 219--224. IEEE, 2021.

\bibitem{schlichtkrull2018modeling}
Michael Schlichtkrull, Thomas~N Kipf, Peter Bloem, Rianne van~den Berg, Ivan
  Titov, and Max Welling.
\newblock Modeling relational data with graph convolutional networks.
\newblock In {\em European semantic web conference}, pages 593--607. Springer,
  2018.

\bibitem{shih2006network}
Hsin-Yu Shih.
\newblock Network characteristics of drive tourism destinations: An application
  of network analysis in tourism.
\newblock {\em Tourism Management}, 27(5):1029--1039, 2006.

\bibitem{simini2021deep}
Filippo Simini, Gianni Barlacchi, Massimilano Luca, and Luca Pappalardo.
\newblock A deep gravity model for mobility flows generation.
\newblock {\em Nature communications}, 12(1):1--13, 2021.

\bibitem{socher2013reasoning}
Richard Socher, Danqi Chen, Christopher~D Manning, and Andrew Ng.
\newblock Reasoning with neural tensor networks for knowledge base completion.
\newblock {\em Advances in neural information processing systems}, 26, 2013.

\bibitem{wang2021multi}
Wei Wang, Junyang Chen, Yushu Zhang, Zhiguo Gong, Neeraj Kumar, and Wei Wei.
\newblock A multi-graph convolutional network framework for tourist flow
  prediction.
\newblock {\em ACM Transactions on Internet Technology (TOIT)}, 21(4):1--13,
  2021.

\bibitem{wang2021spatial}
Yuewei Wang, Hang Chen, and Xinyang Wu.
\newblock Spatial structure characteristics of tourist attraction cooperation
  networks in the yangtze river delta based on tourism flow.
\newblock {\em Sustainability}, 13(21):12036, 2021.

\bibitem{wu2014intra}
Lun Wu, Ye~Zhi, Zhengwei Sui, and Yu~Liu.
\newblock Intra-urban human mobility and activity transition: Evidence from
  social media check-in data.
\newblock {\em PloS one}, 9(5):e97010, 2014.

\bibitem{xu2021characterizing}
Yang Xu, Jingyan Li, Alexander Belyi, and Sangwon Park.
\newblock Characterizing destination networks through mobility traces of
  international tourists—a case study using a nationwide mobile positioning
  dataset.
\newblock {\em Tourism Management}, 82:104195, 2021.

\bibitem{xu2022understanding}
Yang Xu, Dan Zou, Sangwon Park, Qiuping Li, Suhong Zhou, and Xinyu Li.
\newblock Understanding the movement predictability of international travelers
  using a nationwide mobile phone dataset collected in south korea.
\newblock {\em Computers, Environment and Urban Systems}, 92:101753, 2022.

\bibitem{yang2014embedding}
Bishan Yang, Wen-tau Yih, Xiaodong He, Jianfeng Gao, and Li~Deng.
\newblock Embedding entities and relations for learning and inference in
  knowledge bases.
\newblock {\em arXiv preprint arXiv:1412.6575}, 2014.

\bibitem{yao2020spatial}
Xin Yao, Yong Gao, Di~Zhu, Ed~Manley, Jiaoe Wang, and Yu~Liu.
\newblock Spatial origin-destination flow imputation using graph convolutional
  networks.
\newblock {\em IEEE Transactions on Intelligent Transportation Systems},
  22(12):7474--7484, 2020.

\bibitem{yin2022convgcn}
Ganmin Yin, Zhou Huang, Yi~Bao, Han Wang, Linna Li, Xiaolei Ma, and Yi~Zhang.
\newblock Convgcn-rf: A hybrid learning model for commuting flow prediction
  considering geographical semantics and neighborhood effects.
\newblock {\em GeoInformatica}, pages 1--21, 2022.

\bibitem{yuan2022explainability}
Hao Yuan, Haiyang Yu, Shurui Gui, and Shuiwang Ji.
\newblock Explainability in graph neural networks: A taxonomic survey.
\newblock {\em IEEE Transactions on Pattern Analysis and Machine Intelligence},
  2022.

\bibitem{zheng2008geolife}
Yu~Zheng, Longhao Wang, Ruochi Zhang, Xing Xie, and Wei-Ying Ma.
\newblock Geolife: Managing and understanding your past life over maps.
\newblock In {\em The Ninth International Conference on Mobile Data Management
  (mdm 2008)}, pages 211--212. IEEE, 2008.

\bibitem{zipf1946p}
George~Kingsley Zipf.
\newblock The p 1 p 2/d hypothesis: on the intercity movement of persons.
\newblock {\em American sociological review}, 11(6):677--686, 1946.

\end{thebibliography}
\end{document}